\documentstyle[fleqn,10pt,epsfig]{article}
\def\photonatomright{\begin{picture}(3,1.5)(0,0)
                                \put(0,-0.75){\tencircw \symbol{2}}
                                \put(1.5,-0.75){\tencircw \symbol{1}}
                                \put(1.5,0.75){\tencircw \symbol{3}}
                                \put(3,0.75){\tencircw \symbol{0}}
                      \end{picture}
                     }

\def\photonatomup{\begin{picture}(1.5,3)(0,0)
                             \put(-0.75,3){\tencircw \symbol{3}}
                             \put(-0.75,1.5){\tencircw \symbol{2}}
                             \put(0.75,1.5){\tencircw \symbol{0}}
                             \put(0.75,0){\tencircw \symbol{1}}
                   \end{picture}
                  }

\def\photonright{\begin{picture}(30,1.5)(0,0)
                     \multiput(0,0)(3,0){10}{\photonatomright}
                  \end{picture}
                 }

\def\photonrighthalf{\begin{picture}(30,1.5)(0,0)
                     \multiput(0,0)(3,0){5}{\photonatomright}
                  \end{picture}
                 }

\def\photonup{\begin{picture}(1.5,30)(0,0)
                  \multiput(0,0)(0,3){10}{\photonatomup}
               \end{picture}
              }
\def\photonuphalf{\begin{picture}(1.5,15)(0,0)
                      \multiput(0,0)(0,3){5}{\photonatomup}
                   \end{picture}
                  }

\def\fermionup{\begin{picture}(1,30)(0,0)
                     \put(0,0){\vector(0,1){15}}
                     \put(0,15){\line(0,1){15}}
               \end{picture}
              }
\def\fermionuphalf{\begin{picture}(1,15)(0,0)
                         \put(0,0){\vector(0,1){7.5}}
                         \put(0,7.5){\line(0,1){7.5}}
                   \end{picture}
                  }

\def\fermionullhalf{\begin{picture}(15,7.5)(0,0)
                        \put(0,0){\vector(-2,1){7.5}}
                        \put(-7.5,3.75){\line(-2,1){7.5}}
                  \end{picture}
                 }

\def\fermionurrhalf{\begin{picture}(15,7.5)(0,0)
                        \put(-15,-7.5){\vector(2,1){7.5}}
                        \put(-7.5,-3.75){\line(2,1){7.5}}
                  \end{picture}
                 }

\newenvironment{Feynman}[3]{\begin{center}
                            \setlength{\unitlength}{#3 mm}
                            \begin{picture}(#1)(#2)
                            \thicklines
                           }{\end{picture} \end{center}}
\newcommand {\oalf} {\mbox{${\cal O}(\alpha)$}}
\newcommand{\rs}{$\sqrt{s}$}
\newcommand{\epl}{$e^+$}
\newcommand{\emi}{$e^-$}
\newcommand{\ee}{$e^+ e^-$}
\newcommand{\nl}{\nonumber \\}
\newcommand{\bq}{\begin{equation}}
\newcommand{\eq}{\end{equation}}
\newcommand{\ba}{\begin{eqnarray}}
\newcommand{\ea}{\end{eqnarray}}

\newcommand{\zz}{$Z^0$}

\newcommand{\ltwo}{LEP~2~}
\newcommand{\RS}{\sqrt{s}}

\newcommand{\EE}{e^+ e^-}
\newcommand{\ZZ}{Z^0}

\newcommand{\para}{\par\noindent}
\begin{document}
\vspace*{-3cm}
\noindent
{\large{\tt DESY 94-216 \\
            November 1994 }}
\vspace{1.5cm}
%
\begin{center}
{\large \bf COMPLETE INITIAL STATE RADIATION TO OFF-SHELL \zz~PAIR
  PRODUCTION IN \ee~ANNIHILATION\footnote[2]{\noindent Contribution to
    the $IX^{th}$ International Workshop "High Energy Physics and
    Quantum Field Theory", Zvenigorod, Moscow Region, Russia,
    September $15^{th}-22^{nd}~1994$.}}\\
%
\vspace{7mm}
 Dima~Bardin\footnote[3]{On leave of absence from Theor.
   Phys. Lab., JINR, ul. Joliot-Curie 6, RU-141980 Dubna, Moscow
   Region, Russia.},
 Dietrich~Lehner, Tord~Riemann \\
 Deutsches Elektronen-Synchrotron DESY \\
 Institut f\"ur Hochenergiephysik IfH, Zeuthen \\
 Platanenallee 6, D-15738 Zeuthen, Germany
\end{center}
\vspace{3mm}
%
%
\begin{abstract}
\noindent
A cross-section calculation for the Standard Model reaction $\EE
\rightarrow (\ZZ\ZZ) \rightarrow f_1\bar{f_1}f_2\bar{f_2}$ including
the effects of the finite
\zz~width and initial state radiative corrections is presented. The
angular phase space integrations are performed analytically, leaving
the invariant masses for numerical integration. Semi-analytical and
numerical results in the energy range $\RS=150\;GeV$ to $1\;TeV$ are
reported.
\end{abstract}
%
%
%
\section{Introduction}
At \ltwo energies and above \ee~annihilation into four fermions is a
major issue. Monte Carlo approaches to four-fermion production with
and without
inclusion of radiative corrections have been developed by several
authors~\cite{kleiss85,glover90,boos94,pittau94}. Complementing these
results we follow a program of `semi-analytical'
calculations~\cite{wwqed93,4fteup94,zhteup94,zzkazi94,NCbardin94}.
These calculations comprise Initial State QED
corrections~\cite{wwqed93,4fteup94,zzkazi94} and Born computations of
large sets of four-fermion Feynman diagrams~\cite{NCbardin94}. In
this paper we are going to present the semi-analytical results for the
complete \oalf~QED Initial State Radiation (ISR) to \zz~Pair
production in \ee~annihilation, also including finite \zz~width
effects:
\bq
  \EE \rightarrow (\ZZ\ZZ) \rightarrow f_1\bar{f_1}f_2\bar{f_2}(\gamma)~,
  ~~~~~~~~~~~~~~~~~~~~~ f_1\!\neq\!f_2~,~~~f_i\!\neq\!e~~.
  \label{eezz4f}
\eq
Semi-analytical means that all angular phase space
variables are integrated analytically, leaving three invariant masses
for numerical integration. Quasi-experimental cuts on the latter can
be easily implemented.
\para
On-shell \zz~pair production has been discussed long
ago~\cite{Brown78}. Numerical calculations including all
\oalf~electroweak corrections except hard photon brems\-strah\-lung
were reported in~\cite{Denner88} for on-shell and in~\cite{Denner90}
for off-shell \zz~bosons. Below, ISR will be treated completely,
including hard bremssstrahlung corrections.
\para
The paper is organized as follows. In section 2 semi-analytical
results for the off-shell Born cross-section are presented, followed
by the ISR results in section 3. The paper closes in section 4 with a
summary, an outlook and some conclusions.
%
%
\section{The Born Cross-Section}
\begin{figure*}
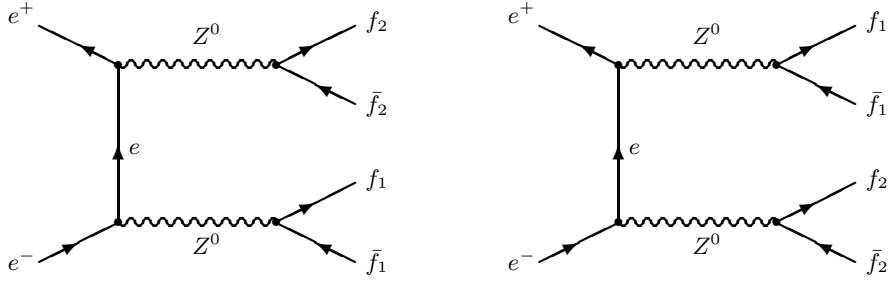

\vspace*{.2cm}
\begin{minipage}[tbh]{7.8cm}{
\begin{center}
\begin{Feynman}{75,60}{-20,0}{0.7}
%
\put(5,15){\fermionurrhalf}
\put(5,45){\fermionullhalf}
\put(5,15){\fermionup}
\put(5,45){\photonright}
\put(5,15){\photonright}
\put(5,45){\circle*{1.5}}
\put(5,15){\circle*{1.5}}
\put(35,15){\circle*{1.5}}
\put(35,45){\circle*{1.5}}
\put(50,7.5){\fermionullhalf}
\put(50,22.5){\fermionurrhalf}
\put(50,37.5){\fermionullhalf}
\put(50,52.5){\fermionurrhalf}
\small
\put(7,28){$e$}
\put(-16,6){\emi}
\put(-16,53){\epl}
\put(19,49){\zz}  
\put(19,09){\zz}
\put(52,06){${\bar f}_1$} 
\put(52,22){$ f_1$}
\put(52,36){${\bar f}_2$}
\put(52,52){$ f_2$}
\normalsize
\put(100,15){\fermionurrhalf}
\put(100,45){\fermionullhalf}
\put(100,15){\fermionup}
\put(100,45){\photonright}
\put(100,15){\photonright}
\put(100,45){\circle*{1.5}}
\put(100,15){\circle*{1.5}}
\put(130,15){\circle*{1.5}}
\put(130,45){\circle*{1.5}}
\put(145,7.5){\fermionullhalf}
\put(145,22.5){\fermionurrhalf}
\put(145,37.5){\fermionullhalf}
\put(145,52.5){\fermionurrhalf}
\small
\put(102,28){$e$}
\put(79,6){\emi}
\put(79,53){\epl}
\put(114,49){\zz}  
\put(114,09){\zz}
\put(147,06){${\bar f}_2$} 
\put(147,22){$ f_2$}
\put(147,36){${\bar f}_1$}
\put(147,52){$ f_1$}
\normalsize
\end{Feynman}
\end{center}
}\end{minipage}
\vspace{-.5cm}
\caption{\it The Born level Feynman diagrams for off-shell \zz~pair
         production, the socalled `{\tt crab}' (or `conversion') diagrams.
         Left: t-channel. Right: u-channel.
        }
\label{zzfeyn}
\end{figure*}

At Born level, process~(\ref{eezz4f}) is described by the two Feynman
diagrams depicted in figure~\ref{zzfeyn}.
%
%
The Born cross-section for process~(\ref{eezz4f}) is given by a simple
double convolution formula
\ba
\sigma^{ZZ}(s) & = &
\int\limits_{4m_1^2}^{(\sqrt{s} - 2m_2)^2} ds_1 \, \rho_Z(s_1)
\int\limits_{4m_2^2}^{(\sqrt{s} - \sqrt{s_1})^2} ds_2 \, \rho_Z(s_2)
\cdot \sigma^{ZZ}_4(s;s_1,s_2) \times 2 \cdot B\!R(1) \cdot B\!R(2)
\label{sigzz}
\ea
invoking the branching ratios $B\!R(1)$~and $B\!R(2)$ for the decay
of the \zz~boson into $f_1\bar{f_1}$~and $f_2\bar{f_2}$ respectively,
the invariant \zz~masses $s_1$ and $s_2$ and
Breit-Wigner density functions for the s-channel
\zz~propagators\footnote[4]{Furtheron we neglect fermion masses
wherever sensible and acceptable for numerical stability.}
\ba
\rho_Z(s_i)
=
\frac{1}{\pi}
\frac {\sqrt{s_i} \, \Gamma_Z (s_i)}
      {|s_i - M_Z^2 + i \sqrt{s_i} \, \Gamma_Z (s_i) |^2}~~~~
      -\!\!\!-\!\!\!-\!\!\!-\!\!\!\!
      \longrightarrow_{\hspace{-1cm}_{\Gamma_Z \rightarrow 0}}
      ~~~~\delta (s_i - M_Z^2).
\label{rhoz}
\ea
The \zz~width is given by
\bq
\Gamma_Z (s_i) =
  \frac{G_{\mu}\, M_Z^2} {24\pi \sqrt{2}} \sqrt{s_i}
  \sum_f (v_f^2+a_f^2).
\label{gzoff}
\eq
Using $a_e\!\!=\!\!1, \,v_e\!\!=\!\!1\!-\!4\sin^2 \!\theta_W,
\,L_e\!\!=\!\!(a_e\!+\!v_e)/2$ and $R_e\!\!=\!\!(a_e\!-\!v_e)/2\;$,
$\;\sigma^{ZZ}_4(s;s_1,s_2)$ is obtained after fivefold
analytical integration over the angular phase space variables:
\ba
\sigma^{ZZ}_4(s;s_1,s_2)
& = & \frac{\lambda^{1/2}}{s}
{\displaystyle \frac{\left(G_{\mu} M_Z^2 \right)^2}{8\pi s}} \!
\left(L_e^4+R_e^4\right) {\cal G}_4^{t+u}(s;s_1,s_2).
\label{sigzz4}
\ea
Algebraic manipulations were
carried out with the help of {\tt FORM}~\cite{form91}.
The sub-index 4 indicates that the underlying matrix element squared
contains four resonant pro\-pa\-ga\-tors.
If considered separately, the contributions
${\cal G}_4^{t}$~from the t-channel, ${\cal G}_4^{u}$~from the
u-channel, and ${\cal G}_4^{tu}$~from the t-u
interference violate unitarity. Due to
so-called unitarity cancellations, their sum ${\cal G}_4^{t+u}$
exhibits proper unitarity behavior and can be very compactly written
as
\ba
   {\cal G}_4^{t+u}(s;s_1,s_2)
   = \frac{s^2+(s_1+s_2)^2}{s-s_1-s_2}{\cal L}_4 - 2~~.
\label{zzmuta}
\ea
This formula was derived in~\cite{4fteup94} and agrees with earlier
results~\cite{Brown78,baier66,cvetic92}. We use
$\lambda \equiv s^2 + s_1^2 + s_2^2 - 2ss_1 - 2 s_1s_2 - 2 s_2s$
{}~~and
\bq
{\cal L}_4 = {\cal L}(s;s_1,s_2) =
\frac{1}{\sqrt{\lambda}} \, \ln \frac{s-s_1-s_2+\sqrt{\lambda}}
                                     {s-s_1-s_2-\sqrt{\lambda}}~.
\label{L4}
\eq
The effect of the finite
\zz~width can be seen from figure~\ref{zzxsec1} as the characteristic
smearing of the peak.
%
\begin{figure}[htb]
\vspace{-2.3cm}
%
\vspace{14.5cm}
\vspace{-.85cm}
\caption[\zz total cross-section 1.]
{\it The total cross-section $\sigma^{ZZ}_{tot}(s)$ for
  process~(\ref{eezz4f}). The numerical input for the figures is:
  $\alpha = 1/137.0359895,~ G_{\mu} = 1.16639\times10^{-5}~GeV^{-2},
   ~m_e = 0.51099906~MeV,~M_Z = 91.173~GeV,~\Gamma_Z = 2.487~GeV,
   ~\sin^2\theta = 0.2325.$
  The numerical precision is estimated to be better than 0.01 permille.}
\label{zzxsec1}
\end{figure}
It should be mentioned that gauge violation problems arise with the
introduction of finite boson widths. It is, however, generally agreed
that these are not critical for process~(\ref{eezz4f}). We only
peripherically mention that a scheme to avoid these gauge violations
was proposed, but
gives incorrect results around threshold~\cite{wwgauge}. Up to now no
treatment of finite widths in boson pair processes seems theoretically
satisfactory.
\para
In order to prevent potential confusion we mention that the
convention used in~\cite{Denner90} has been adopted, namely to present
the total cross-section
$\sigma^{ZZ}_{tot} \equiv \sigma^{ZZ}/ (2 \cdot B\!R(1) \cdot B\!R(2))$
in the plots.
This is adequate in the sense that in the narrow width approximation
$\sigma^{ZZ}_{tot}$
recovers the cross-section for on-shell \zz~pair production which is
seen if a summation over the decay channels $f_i\bar{f_i}$ of
the \zz~bosons is performed. Using the result for on-shell \zz~pair
production $\sigma^{ZZ}_{on-shell} =
\sigma^{ZZ}_4(s;s_1,s_2)$~\cite{zzkazi94,Brown78} we obtain:
\ba
   \sigma^{ZZ}_{on-shell}
   & \equiv & \sigma^{ZZ}_{on-shell} \times 1^2
   \;\;  = \;\;  \sigma^{ZZ}_{on-shell} \left[ \sum_i BR(i) \right]^2
   \nonumber \\
   & = & \sigma^{ZZ}_{on-shell}
           \left[
             \sum_i BR(i)^2 + \sum_{i<j}2 \cdot BR(i)\cdot BR(j)
           \right]
   \nonumber \\
   & \approx &  \sum_{i\leq j} \sigma^{ZZ}_{narrow~\!width}
\ea
where $i\leq j$ in the summation over the narrow width
four-fermion cross-section is necessary to avoid double counting.
For $i\neq j$, the terms $\sigma^{ZZ}_{narrow~\!width}$~are given by
the narrow width approximation of the right hand side of
eq.~(\ref{sigzz}).
This convention of ours was not emphasized
in earlier publications~\cite{4fteup94,zzkazi94}.
%
%
\section{${\cal O}(\alpha)$ Initial State Radiation}
In \ee~annihilation, ISR is known to represent the bulk of the
radiative corrections.
The ${\cal O}(\alpha)$ `amputated' Feynman diagrams for Initial State
Bremsstrahlung (ISB) to process~(\ref{eezz4f}) are shown in
figure~\ref{zzbrem}. The corresponding virtual ISR diagrams are given
in figure~\ref{zzvirt}. External leg self energies are absorbed into
the on-shell renormalization.
%
\begin{figure}[bt]
\begin{minipage}[tbh]{15.cm} {
\begin{center}
\begin{Feynman}{150,60}{-43,0}{0.7}
\small
\put(-26,52.5){$e^+$}  
\put(-26,06.5){$e^-$}
\put(5,49){$\gamma(p)$}
\normalsize
\put(-5,45){\line(-2,1){15.00}}
\put(-16.25,50.625){\vector(-2,1){1}}  
\put(-8.75,46.875){\vector(-2,1){1}}
\put(-5,15){\fermionurrhalf}
\put(-5,15){\fermionup}
\put(-12.5,48.75){\photonrighthalf}
\put(-5,45){\circle*{1.5}}
\put(-5,15){\circle*{1.5}}
\put(-12.5,48.75){\circle*{1.5}}
%
\put(85,45){\fermionullhalf}
\put(85,15){\fermionup}
\put(76.5,10.75){\photonrighthalf}
\put(85,45){\circle*{1.5}}
\put(85,15){\circle*{1.5}}
\put(76.5,10.75){\circle*{1.5}}
\put(70,7.5){\line(2,1){15.00}}
\put(73.75,9.375){\vector(2,1){1}}  
\put(83,14){\vector(2,1){1}}
%
\put(40,15){\fermionurrhalf}
\put(40,45){\fermionullhalf}
\put(40,30){\fermionuphalf}
\put(40,15){\fermionuphalf}
\put(40,30){\photonrighthalf}
\put(40,45){\circle*{1.5}}
\put(40,15){\circle*{1.5}}
\put(40,30){\circle*{1.5}}
\end{Feynman}
\end{center}
}\end{minipage}
\vspace{-.5cm}
\caption{\it The amputated ISB diagrams for \zz~pair production.}
\label{zzbrem}
%
%
\vspace{.3cm}
\begin{minipage}[tbh]{15cm} {
\begin{center}
\begin{Feynman}{150,60}{-44,0}{0.78}
\small
\put(-42,52){$e^+$}  
\put(-42,6){$e^-$}
\put(-21,15){\fermionurrhalf}
\put(-21,45){\fermionullhalf}
\put(-21,15){\fermionuphalf}
\put(-21,30){\photonuphalf}
\put(-21,37.5){\oval(15,15)[r]}
\put(-15.6,42.8){\vector(-1,1){1}}  
\put(-15.55,32.2){\vector(1,1){1}}  
\put(-21,45){\circle*{1.5}}
\put(-21,15){\circle*{1.5}}
\put(-21,30){\circle*{1.5}}
\put(-13.5,37.5){\circle*{1.5}}
\put(16,15){\fermionurrhalf}
\put(16,45){\fermionullhalf}
\put(16,30){\fermionuphalf}
\put(16,15){\photonuphalf}
\put(16,22.5){\oval(15,15)[r]}
\put(21.35,27.8){\vector(-1,1){1}}  
\put(21.35,17.2){\vector(1,1){1}}  
\put(16,45){\circle*{1.5}}
\put(16,15){\circle*{1.5}}
\put(16,30){\circle*{1.5}}
\put(23.50,22.5){\circle*{1.5}}
\put(53,15){\fermionurrhalf}
\put(53,45){\fermionullhalf}
\put(53,15){\line(0,1){7.5}}
\put(53,18.75){\vector(0,1){1}}  
\put(53,41.25){\vector(0,1){1}}  
\put(53,37.5){\line(0,1){7.5}}
\put(53,22.5){\photonuphalf}
\put(53,30){\oval(15,15)[r]}
\put(60.48,30){\vector(0,1){1}}  
\put(53,45){\circle*{1.5}}
\put(53,15){\circle*{1.5}}
\put(53,37.5){\circle*{1.5}}
\put(90,15){\fermionurrhalf}
\put(105,22.5){\fermionurrhalf}
\put(90,45){\fermionullhalf}
\put(105,37.5){\fermionullhalf}
\put(105,22.5){\fermionuphalf}
\put(90,15){\photonup}
\put(90,45){\circle*{1.5}}
\put(90,15){\circle*{1.5}}
\put(105,22.5){\circle*{1.5}}
\put(105,37.5){\circle*{1.5}}
\end{Feynman}
\end{center}
}\end{minipage}
\vspace{-.6cm}
\caption{\it The amputated virtual ISR diagrams for \zz~pair
         production.}
\vspace{.3cm}
\label{zzvirt}
\end{figure}
The double-differential cross-section for off-shell \zz~pair
production including ${\cal O}(\alpha)$ ISR with soft photon
exponentiation can be presented as
\bq
  \frac{d^2 \sigma^{ZZ}}{ds_1 ds_2}
  =
  \int\limits_{(\sqrt{s_1}+\sqrt{s_2})^2}^s \!\!\!\!
  \frac{ds'}{s}
  \, \rho(s_1) \, \rho(s_2) \,
  \left[ \beta_e v^{\beta_e - 1} {\cal S} + {\cal H} \right]
  \label{compqed}
\eq
with $\beta_e \!=\!2 \alpha/\pi [ \ln (s/m_e^2) - 1 ]$ and $v=1-s'/s$.
The soft+virtual and hard photonic parts ${\cal S}$ and ${\cal H}$
are calculated analytically, requiring seven angular integrations.
Both separate into a universal part with the Born cross-section
factorizing and a nonuniversal part:
\ba
  {\cal S}(s,s';s_1,s_2) & = &
  \left[1 + {\bar S}(s) \right] \sigma_0(s';s_1,s_2)
  + \;\sigma_{\hat S}(s';s_1,s_2) ~~~~,
  \nl
  \hspace{-1cm}
  {\cal H}(s,s';s_1,s_2) & = &
  \underbrace{{\bar H}(s,s') ~\sigma_0(s';s_1,s_2)~~~}_{Universal\;
    Part}
  + \underbrace{\sigma_{\hat H}(s,s';s_1,s_2)}_{Nonuniversal\;Part}.
\ea
An explicit derivation proved that ${\bar S}$ and
${\bar H}$ are identical to the radiators known
from s-channel fermion pair production~\cite{4fteup94,zzkazi94,radiators}:
\ba
  {\bar S}(s) = \frac{\alpha}{\pi}
                \left[  \frac{\pi^2}{3} - \frac{1}{2} \right]
                + \frac{3}{4}\beta_e + {\cal O}(\alpha^2)~,
  \nl
  {\bar H}(s,s') = - \frac{1}{2}
                     \left(1+\frac{s'}{s}\right)\beta_e
                   + {\cal O}(\alpha^2)~.
\ea
The analytical formulae for the non\-uni\-ver\-sal
con\-tri\-bu\-tions are very involved as they contain many Dilogarithm
and Trilogarithm functions and will therefore be published
elsewhere. Compared to the universal corrections, nonuniversal
corrections are small, because they do not contain the mass
singularity $\beta_e$. It is seen from the numerical results
presented in figure~\ref{zzxsec2} that the relative contribution of
nonuniversal corrections is negligible around threshold and inreases
to 3\% at 1~TeV.
%
\begin{figure}[bt]
\vspace{-1.5cm}
%
\vspace{14.5cm}
\vspace{-.85cm}
\caption[\zz total cross-section 2.]
{\it The effect of universal and non\-uni\-ver\-sal ISR on
  process~(\ref{eezz4f}) for off-shell \zz~bosons. The inlay in the
  upper right corner exhibits the \rs-dependence of the ratio of
  the universally (dash-dotted line) and non\-uni\-ver\-sally (dotted
  line) ISR corrected over the Born total cross-section. The numerical
  precision is estimated to be better than 0.1 permille.}
\label{zzxsec2}
\end{figure}
%
As in the nonuniversal corrections to W-pair production we observe the
so-called screening property, i.e. the nonuniversal corrections are
damped by a factor $s_1s_2/s^2$~\cite{wwqed93}.
It is seen from both figure~\ref{zzxsec1} and figure~\ref{zzxsec2}
that ISR plays an important r\^{o}le in
almost the whole energy range under consideration. Around threshold,
in the energy range of \ltwo, a peak in the ISR corrections' relative
importance can be seen.
%
%
\section{Summary, Outlook and Conclusions}
We have presented finite width and initial state QED corrections to
$\EE \!\!\rightarrow \!(\ZZ\ZZ) \!\rightarrow f_1\bar{f_1}f_2\bar{f_2}$
in a semi-analytical approach. It was shown that both yield important
corrections to the total cross-section.
\para
The inclusion of photons replacing \zz~bosons in the
{\tt crab} Feynman diagrams of
figure~\ref{zzfeyn} is necessary to endow our computations with
experimental relevance. This inclusion is straightforward, but will
also require a refined treatment of final state fermion masses. In a
second step it is intended to merge the results of the
multi-diagram, semi-analytical Born calculation given in
reference~\cite{NCbardin94} with ISR calculations as performed above
to yield a precise result for some channels of
the process $\;e^+ e^- \rightarrow l  \bar l \; q  \bar q$, probably
already matching the feasible experimental precision. The set of
Feynman diagrams for such a calculation is given
in~\cite{NCbardin94}. It is noteworthy that the {\tt crab}
diagrams, constitute the by far most important
contribution to the cross-section if \rs~is far above $M_Z$.
An extension of the presented analysis to QCD initial state gluon
radiation in \zz~pair production at LHC is also within reach.
The authors acknowledge discussions with Arnd Leike and Uwe M\"uller.
%
%
%

%
%
\end{document}